# Cost-effective photonic super-resolution millimeter-wave joint radar-communication system using self-coherent detection

**Wenlin Bai,**[1] **Peixuan Li,**[1,*] **Xihua Zou,**[1,*] **Ningyuan Zhong,**[1] **Wei Pan,**[1] **Lianshan Yan,**[1] **and Bin Luo**[1]

[1] *Center for Information Photonics and Communications, School of Information Science and Technology, Southwest Jiaotong University, Chengdu 611756, China*
*\*Corresponding authors: lipeixuan@swjtu.edu.cn; zouxihua@swjtu.edu.cn*

**A cost-effective millimeter-wave (MMW) joint radar-communication (JRC) system with super resolution is proposed and experimentally demonstrated, using optical heterodyne up-conversion and self-coherent detection down-conversion techniques. The point lies in the designed coherent dual-band constant envelope linear frequency modulation-orthogonal frequency division multiplexing (LFM-OFDM) signal with opposite phase modulation indexes for the JRC system. Then the self-coherent detection, as a simple and low-cost means, is accordingly facilitated for both de-chirping of MMW radar and frequency down-conversion reception of MMW communication, which circumvents the costly high-speed mixers along with MMW local oscillators and more significantly achieves the real-time decomposition of radar and communication information. Furthermore, a super resolution radar range profile is realized through the coherent fusion processing of dual-band JRC signal. In experiments, a dual-band LFM-OFDM JRC signal centered at 54-GHz and 61-GHz is generated. The dual bands are featured with an identical instantaneous bandwidth of 2 GHz and carry an OFDM signal of 1 GBaud, which help to achieve a 6-Gbit/s data rate for communication and a 1.76-cm range resolution for radar.** © 2022 Optica Publishing Group

---

Fueled by critical applications of the fifth and sixth generation (5G and 6G) communications heavily rely on accurate sensing and localization (e.g., smart homes, autonomous driving, smart manufacturing, et.al.), joint radar-communication (JRC) technologies are gaining more and more attention [1]. Particularly, as expected in the 3rd-generation partnership project (3GPP) new radio Release 17 [2], the millimeter-wave (MMW) frequency band up to 71-GHz is able to greatly enhance the network capacity and sensing ability. Whilst, the real-world deployment of MMW JRC systems is haunted by challenges of high path-loss/cost/complexity in terms of signal transmission, generation and processing.

Thanks to the distinct advantages of high operating frequency, large instantaneous bandwidth (IBW), low frequency-dependent loss and strong immunity to electromagnetic interferences, microwave photonics technology has been intensively explored to enable advanced JRC systems of high-resolution radar detection and large-capacity wireless transmission [3-9]. The reported photonics-assisted JRC approaches can be classified into three categories: i.e., frequency-division multiplexing [3], time-division multiplexing [4, 5], and joint waveform strategies [6-9]. Using the frequency-division multiplexing scheme, a low-cost dual-band radar and communication system facilitated by a single photonics-assisted transceiver is proposed, having a 54-Mbit/s data rate and 7.5-m radar range resolution [3]. Alternatively, the W-band [4] and THz-band [5] photonic JRC systems in the time-division multiplexing mode are successively proposed and experimentally demonstrated, showing a communication data rate as high as 38.1-Gbit/s and the radar range resolution up to 0.94-cm. Among them, the joint waveform strategy stands out as the most efficient way to explore spectral and temporal resources. As such, in [6], a linear frequency modulation (LFM)-encoded amplitude shift key (ASK) JRC waveform is generated based on the cascaded Mach-Zander modulator (MZM), achieving a 1.8-cm radar range resolution and a ~100-Mbit/s data rate. Subsequently, the orthogonal phase-coded JRC signals based on the photonic spectrum-spreading phase-coding [7] and the optoelectronic oscillator (OEO) [8] are generated for obtaining the higher radar peak-to-sidelobe ratio performance (> 12-dB) and data rate (> 1-Gbit/s). Furthermore, the high-spectral-efficiency orthogonal frequency division multiplexing (OFDM)-based JRC system using the OEO technique reaches a high data rate of 6.4-Gbit/s and a radar range resolution of 7.5-cm [9]. Nonetheless, these efforts either suffer from the issue of low efficiency in temporal/spectral resources [3-5] or fail to offer inspirations for implementing spectrally efficient joint waveform-based JRC systems operating up to the MMW band.

In this letter, we propose and experimentally demonstrate an MMW JRC system using the joint waveform scheme, particularly dedicating efforts to enhance cost-effectiveness and easiness-to-implement. Our proposal lies on coherent dual-band constant envelope LFM-OFDM signals designed by modulating the phase of two up-chirp LFM carriers using two replicas of the OFDM communication signal with opposite phase modulation indexes (PMIs). Accordingly, the self-coherent detection is reaped for MMW radar de-chirping and communication down-conversion reception, which decomposes the radar and communication information in a real-time mode. Thus, it is able to achieve cost-effective radar and communication functions of high-resolution radar detection and large-capacity wireless transmission. Moreover, the coherent fusion processing (CFP) of dual-band JRC signals facilitates a super-resolution radar range profile beyond the IBW limit and therefore relaxes the processing speed requirements of electronic devices. In experiments, dual-band coherent MMW JRC signals centered at 54-GHz and 61-GHz are generated by optical heterodyne beating, having the identical 2 GHz IBW for LFM signal and 1 Gbaud symbol rate for 64-quadrature amplitude modulation (64-QAM) OFDM signal. With the aid of cost-effective self-coherent detection and advanced CFP, our proposal experimentally shows a communication data rate of 6-Gbit/s and a radar detection resolution equivalent to that from a 9-GHz IBW LFM signal, 1.76-cm.

Figure 1 shows the schematic diagram of our proposed JRC system. The designed intermediate frequency (IF) coherent dual-band constant-envelop (CE)-LFM-OFDM JRC signal can be formulated as

$$s_{IF-JRC}(t) = A\{\exp j[\omega_1 t + \pi k t^2 + 2\pi h m(t)] \\ + \exp j[\omega_2 t + \pi k t^2 - 2\pi h m(t)]\}, \quad (1)$$

where $A$ is the signal amplitude, and $\omega_1, \omega_2$ are the initial angular frequencies of two LFM carriers, respectively. $k = B/T_c$ denotes the slope of the LFM carrier, which is related to its IBW $B$ and pulse width $T_c$. $h$ represents the PMI. $m(t)$ is the real-value OFDM baseband signal. It is noted that the IF JRC signals of distinct bands (around $\omega_1$ and $\omega_2$) have opposite PMIs for facilitating self-coherent detection.

As shown in Fig. 1, the optical carrier from a continuous-wave laser is injected into the transmitter MZM (MZM1). This modulator is driven by a single-tone radio frequency (RF) signal, and biased at the minimum transmission point to obtain a carrier-suppressed double sideband (CS-DSB) optical signal. Under the small signal modulation assumption, the output electrical field of MZM1 can be described as

$$E_{MZM1}(t) \propto E_1 J_1(\beta) \exp j\omega_c t \{\exp -j\omega_{RF} t + \exp j\omega_{RF} t\}, \quad (2)$$

where $E_1, \omega_c$ are the amplitude and center angular frequency of the optical carrier. $\omega_{RF}$ is the center angular frequency of the driving RF signal. $J_n(\cdot)$ designates the $n$-th order Bessel function of the first kind, and $\beta$ stands for the modulation index of MZM1.

Then, the CS-DSB optical signal is sent to an optical inter-leaver to separate two optical sidebands. The optical sideband in the upper path serves as the data-carrying optical carrier to be injected into a dual-parallel MZM (DPMZM). According to [10], the IF JRC signal generated by the arbitrary waveform generator (AWG) passes through an electrical 90º hybrid to drive the DPMZM for achieving a carrier-suppressed single sideband (CS-SSB) modulation. The state of polarization and power of the optical sideband in the lower path are adjusted through the polarization controller and variable optical attenuator to be aligned with those of the CS-SSB optical signal in the upper path. After that, the optical signals in two paths are combined via an optical coupler (OC), which can be expressed by

$$E_{OC}(t) \propto E_1 J_1(\beta) \exp j[(\omega_c - \omega_{RF})t] + E_1 J_1(\beta) E_2 J_1(\gamma) \cdot \\ \begin{cases} \exp j[(\omega_c + \omega_{RF} + \omega_1)t + \pi k t^2 + 2\pi h m(t)] \\ + \exp j[(\omega_c + \omega_{RF} + \omega_2)t + \pi k t^2 - 2\pi h m(t)] \end{cases}, \quad (3)$$

where $E_2$ is the amplitude of the CS-SSB optical signal, $\gamma$ denotes the modulation index of DPMZM. The output of OC is then divided into two optical paths. One serves as the reference signal for the self-coherent radar de-chirping process. Another one is boosted by an Erbium-doped fiber amplifier (EDFA) and sent to the transmitter photodetector (PD1) to generate MMW JRC signals. With the aid of optical-to-electrical conversion, the recovered photocurrent can be written as

$$i_{PD1}(t) \propto R\{\cos[(2\omega_{RF} + \omega_1)t + \pi k t^2 + 2\pi h m(t)] \\ + \cos[(2\omega_{RF} + \omega_2)t + \pi k t^2 - 2\pi h m(t)]\}, \quad (4)$$

where $R$ denotes the amplitude of photocurrent. Here, the DC and low-frequency terms are ignored. Therefore, the coherent dual-band MMW JRC signals centered at $2\omega_{RF} + \omega_1$ and $2\omega_{RF} + \omega_2$ are generated by optical heterodyne beating detection.

Subsequently, the generated MMW JRC signal is radiated into free space through the power amplifier (PA) and antenna. At the communication receiver, the self-coherent square-law envelope detection (ED) is used to down-convert the MMW JRC signal to an IF one:

$$i_{ED}(t) \propto G\{\cos[(\omega_2 - \omega_1)t + 4\pi h m(t)] + \cos[(\omega_2 - \omega_1)t]\}, \quad (5)$$

where $G$ represents the amplitude of the IF signal. Thus, with the self-coherent detection technique, the expensive and complicated coherent down-conversion requiring high-speed mixers and MMW local oscillators (LOs) is averted in the communication end. More significantly, see Eq. (5), the interfering radar LFM carrier is directly removed, leaving a single-carrier IF CE-OFDM signal [11]. Thus, it greatly saves further efforts for the decoupling between radar and communication information in a JRC signal. Also, as can be seen in Eq. (5), the PMI of the decomposed CE-OFDM signal is twice that in the transmit IF JRC signal, which is beneficial for improving the bit error ratio (BER) performance of the system [11].

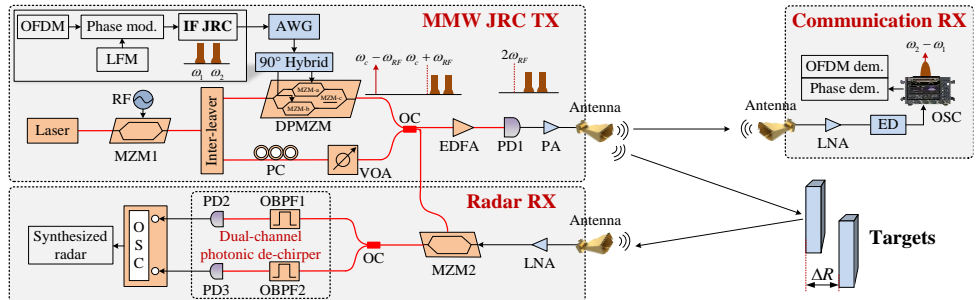

**Fig. 1.** Schematic diagram of proposed photonic MMW JRC system. AWG: arbitrary waveform generator; ED: envelope detector; EDFA: Erbium-doped fiber amplifier; IF JRC: immediate frequency joint radar and communication; LFM: linear frequency modulation; LNA: low noise amplifier; MZM: Mach-Zehnder modulator; DPMZM: dual-parallel MZM; OBPF: optical band-pass filter; OC: optical coupler; OSC: oscilloscope; PA: power amplifier; PC: polarization controller; PD: photodetector; RX: receiver; RF: radio frequency; TX: transmitter; VOA: variable optical attenuator.

On the other hand, using the well-stablished microwave photonics de-chirping scheme [10], LO-free dual-parallel self-coherent de-chirpings for the MMW dual-band radar echoes can also be achieved. As shown in Fig. 1, echoes reflected by targets are received by an MMW antenna and amplified by a low noise amplifier (LNA). The output of LNA is applied to MZM2 to modulate the reference optical signal. Here, MZM2 is biased at the quadrature point. Then, the output of MZM2 is split into two parallel optical de-chirping channels both entailing an optical band-pass filter (OBPF) and a low-speed PD, corresponding to two MMW echoes of different bands. The $i-th$ ($i=1,2$) OBPF at one de-chirping channel is used to select specific optical signals around the sideband induced by one sub-band of MMW echoes, of which the output can be written as

$$E_{OBPF\_i} \propto \{\exp j[(\omega_c + \omega_{RF} + \omega_i + \pi kt)t \pm 2\pi hm(t)] \\ + \exp j[(\omega_c + \omega_{RF} + \omega_i + \pi kt + k\tau)t \pm 2\pi hm(t-\tau)]\}, \quad (6)$$

where $\tau$ denotes the round-trip delay of echo with respect to the transmit signal. Then, the selected optical signals are sent to PDs for obtaining radar de-chirped signals of two radar sub-bands based on the square-law detection:

$$i_{de-chirped\_i}(t) \propto C_i \cos[k\tau t \pm 2\pi hm'(t)], \quad (7)$$

where $C_i$ is the amplitude of the de-chirped signal of the $i-th$ sub-band. $m'(t)=m(t)+m(t-\tau)$ contains OFDM signal and its delayed version. Therefore, dual-parallel self-coherent detection-assisted de-chirpings for the dual-band MMW echoes are realized. More importantly, the occupying frequency band of de-chirping signals is generally within DC to tens of MHz. Correspondingly, the communication interferences for radar can be easily remedied by properly setting the communication band away from the low-frequency region, and thus achieving a real-time decomposition of radar detection and communication as well.

Also, as the JRC signals of different sub-bands share a single photonic transceiver, the coherence among the attained de-chirped signals of two sub-bands can also be guaranteed, which averts the complicated phase compensation between them. Subsequently, two de-chirped signals of two sparse sub-bands can be fused by CFP [12] to gain a super-resolution range profile of the target beyond the detection capability of the single-band radar. The CFP here is based on the well-investigated algorithm presented in [12], to realize paddings of spectrum gap with interpolating values for the missing band between two sparse sub-bands. As a results, the corresponding range resolutions for two sub-bands are governed by $\Delta r_1 = c/2B_1, \Delta r_2 = c/2B_2$, respectively, in which $c$ is the velocity of light and $B_1, B_2$ are bandwidths of two sub-bands, respectively. Whilst, the range resolution for the fused data can reach $\Delta R = c/\Delta B$, where $\Delta B = B_1 + B_2 + B_3$, and $B_3$ is the bandwidth of the missing band between two sub-bands.

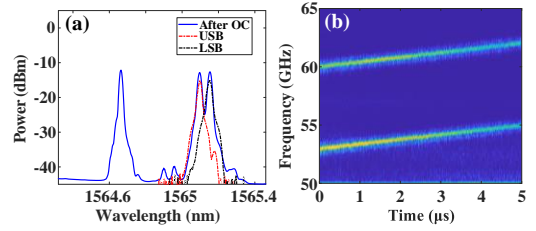

**Fig. 2.** (a) Measured optical spectral after OC (blue line), at the outputs of OBPF1 (red line) and OBPF2 (black line); (b) spectrogram of the generated coherent dual-band MMW JRC signals after PD.

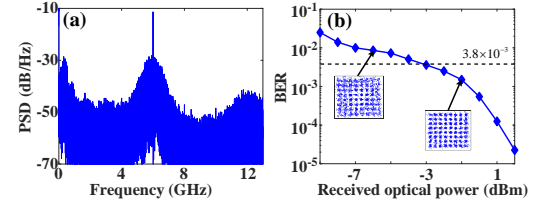

**Fig. 3.** (a) Measured power spectral density of the down-converted single-carrier IF CE-OFDM signal; (b) Dependence of BERs on the received optical power. (Inset: constellation diagrams of the demodulated 64-QAM signal)

Proof-of-concept experiments is carried out based on the setup shown in Fig. 1. The optical carrier from a laser source is with a wavelength of 1564.86-nm and a 12-dB power level. MZM1 has a 3-dB bandwidth of 40-GHz and is driven by a 25-GHz single-tone RF signal. A 50-GHz optical inter-leaver is then used to separate two sidebands of the CS-DSB signal from MZM1. The off-line generated IF JRC signal is loaded into a 64-GSa/s AWG for digital-to-analog conversion. The output of AWG is applied to the DPMZM to modulate the optical sideband in the upper path for CS-SSB modulation. The output optical spectrum of the OC is measured and shown in Fig. 2(a). After being amplified by an EDFA, the output of OC is sent to a 70-GHz PD for generating the MMW JRC signal.

For the transmit IF dual-band JRC signal, two up-chirp LFM carriers centered at 4-GHz and 11-GHz are phase modulated by two OFDM signals with opposite PMIs (0.7 and -0.7). Two LFM bands have identical IBW and pulse width of 2 GHz and 5 μs. The baseband real-valued OFDM communication signal is with a 1 GHz bandwidth and modulated in the 64-QAM format. When such an IF JRC signal is applied to the proposed photonic MMW JRC system, dual-band MMW JRC signals centered at 54-GHz and 61-GHz are generated, of which the time-frequency characteristics are shown in Fig 2(b).

A 1.5-m 60-GHz MMW wireless transmission link for communication function demonstrations is established by a PA and a pair of horn antennas. After wireless transmission, the coherent dual-band MMW JRC signals are captured by the antenna at the communication receiver and amplified by an LNA to compensate for the free-space loss. A low-cost amplitude detector (AD) is used to implement the self-coherent MMW down-conversion reception. The measured power spectral density at the output of AD can be seen in Fig. 3(a), showing a single-carrier IF CE-OFDM signal centered around 7-GHz and only containing communication information. It results from the self-mixing between two sub-bands (at 54-GHz and 61-GHz) of the received dual-band MMW JRC signal, which also leads to the mutual cancellation between two up-chirp LFM carriers of two sub-bands for decomposing radar interferences and communication information. Subsequently, the down-converted IF signal is digitized by an oscilloscope running at a sampling rate of 40 GSa/s. Off-line DSPs only for demodulating the CE-OFDM signal are then employed to evaluate the communication performance of our system. Fig. 3(b) shows the dependence of obtained BER on the received optical power before PD1. From Fig. 3(b), the BER performance of our system can reach the 7 % pre-forward error correction threshold ($3.8\times10^{-3}$) for a raw data rate of 6-Gbit/s. In addition, the optimal BER can be as low as $2.2\times10^{-5}$ for a 2-dBm received optical power, offering a considerable performance margin for longer fiber and wireless transmissions.

For radar function demonstrations, two metal reflectors with a size of 2-cm × 3-cm are used to emulate the targets of interest, as shown in Fig. 4(a). The dual-band MMW JRC signals are radiated into free space by the PA and radar transmit antenna to detect these targets ~2.6-m away from them. Echoes reflected by targets are acquired by a MMW receiving antenna, and applied to MZM2 with a wideband bandwidth of 65-GHz. Through a 50:50 OC, the output of MZM2 is divided into two copies and injected into the dual-channel photonic de-chirper. Two OBPFs are leveraged to separate up-sideband (USB) and low-sideband (LSB) optical signals for the dual-parallel radar de-chirping processes of two sub-bands. The measured optical spectra at the output of two OBPFs is shown in Fig. 2(b). After the optical-to-electrical conversions in low-speed PDs (PD2 and PD3), two radar de-chirped signals are digitized by two channels of an OSC with a low sampling rate of 250-MSa/s. Figs. 4(b) and (c) present the obtained radar range profile results of two reflectors separated by 7.5-cm and 1.67-cm intervals, in which the blue, black, and red lines denote the measured results of the USB, LSB and their fused data, respectively. As shown in Fig. 4(b), the obtained distance between two targets is 7.6-cm, which is consistent with the theoretical 7.5-cm range resolution for the independent USB and LSB signals having a 2 GHz IBW. When two targets separate at 1.67-cm, they cannot be distinguished by a single USB or LSB signal, as seen in Fig. 4(c). However, they can be distinguished by the fused data, and the measured distance is 1.76-cm, which agrees well with the theoretical super-resolution of 1.67-cm for the fused data with a 9 GHz equivalent IBW.

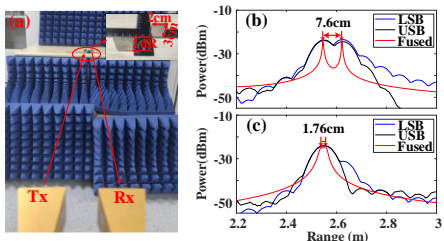

**Fig. 4.** (a)Photograph of two metal targets for radar detection; obtained radar range results of two reflectors separated by (b) 7.5-cm, and (c) 1.67-cm intervals.

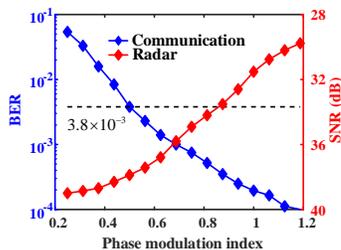

**Fig. 5.** BERs of demodulated communication signal and SNRs of radar de-chirped signal versus the PMI of the transmitted coherent IF JRC signal.

Furthermore, an additional experiment is carried out to evaluate the performance trade-off between radar and communication functions governed by the PMIs of the transmitted IF LFM-OFDM JRC signal. Figure 5 shows the BER of the demodulated communication signal and SNR of the radar de-chirped signal versus different values of PMI ranging from 0.2 to 1.2. One can observe that the suggested value of PMI for our system should be 0.7, which is able to balance communication and radar performances.

In conclusion, we have proposed and experimentally validated a photonics-assisted cost-effective MMW JRC system based on optical heterodyne up-conversion and self-coherent detection down-conversion techniques. The key assets of such a proposal are manifold: firstly, both MMW-band radar and communication receptions are achieved through self-coherent detection; secondly, MMW radar and communication system operating from 53-GHz to 62-GHz is achieved in an MMW-LOs-free mode; thirdly, the self-coherent detection significantly realizes the real-time decomposition of radar and communication information and accomplishes a 6-Gbit/s high-speed wireless communication transmission; finally, using the CFP of dual-band JRC signals, a 1.76-cm super-resolution radar range profile is achieved based on a low-speed (or low sample rate) and narrow-band DSP unit. Therefore, this proposal can be expected to offer a cost-effective photonics-assisted alternative for future 6G and beyond integrated high-resolution sensing and high-capacity communication application scenarios.

**Funding.** National Natural Science Foundation of China (61922069, U21A20507, 62001401), Sichuan Science and Technology Program (2022NSFSC0559, 2022JDTD0013), and Fundamental Research Funds for the Central Universities (2682021CX045).

**Disclosures.** The authors declare no conflicts of interest.

**Data availability.** Data underlying the results presented in this paper are not publicly available at this time but may be obtained from the authors upon reasonable request.